\documentclass[a4paper,12pt]{article} % default is 10 pt
\usepackage[T1]{fontenc}
\usepackage{amsmath,amssymb,amsfonts}        % moduli per le estensioni ed i fonts AMS
\usepackage{mathptmx}
\usepackage[scaled=0.91]{helvet}
\usepackage{courier}
\usepackage[dvips]{graphicx} % needed for including graphics e.g. EPS, PS
\usepackage{psfrag}

\long\def\comment#1{}

\title{An Analytical and Numerical Study of Liquid Dynamics in a 1D Capillary under Entrapped Gas Action}

\author{Riccardo Fazio\footnote{Corresponding author {\tt e-mail: rfazio@dipmat.unime.it}}\ \ and Salvatore Iacono\\[1ex]
Department of Mathematics and Computer Science\\
University of Messina \\ 
Viale F. Stagno D'Alcontres 31, 98166 Messina, Italy}

\date{\small Submitted to Mathematical Methods in the Applied Sciences on September 3, 2012 and in revised form on September 1, 2013}

\begin{document}
\maketitle

% >>>>>>>>>>>>>>>>>>>>>>>>> Keywords and Abstract <<<<<<<<<<<<<<<<<<<<<
\begin{abstract}
Capillary dynamics has been and is yet an important field of research, because of its very relevant role played as the core mechanism at the base of many applications. 
In this context, we are particularly interested in the liquid penetration inspection technique. 
Due to the obviously needed level of reliability involved with such a non-destructive test, this paper is devoted to study how the presence of an entrapped gas in a close-end capillary may affect the inspection outcome. 
This study is carried out through a 1D ordinary differential model that despite its simplicity is able to point out quite well the capillary dynamics under the effect of an entrapped gas. 
The paper is divided into two main parts; the first starts from an introductory historical review of capillary flows modeling, goes on presenting the 1D second order ordinary differential model, taking into account the presence of an entrapped gas and therefore ends by showing some numerical simulation results. 
The second part is devoted to the analytical study of the model by separating the initial transitory behavior from the stationary one. 
Besides, these solutions are compared with the numerical ones and finally an expression is deduced for the threshold radius switching from a fully damped transitory to an oscillatory one.   
\end{abstract}

%\vspace{1cm}



\section{Introduction}\label{Introduction}
Capillarity is a well known physical phenomenon directly related to the free energy present at the interface liquid-air. 
Whenever the liquid gets in touch with a solid capillary, the motion of the air-liquid interface meniscus takes place, according to the wettability of the capillary wall. 
The force responsible for such motion is just the so-called surface tension. 

Laplace and Young were among the first to study the surface tension and capillarity, at the beginning of the nineteenth century. 
In particular, their work started from the awareness that the static pressure on the liquid side of a liquid-air interface is reduced by the effect of the surface tension. 
Some time later, Hagen %\cite{Hagen:1839} 
and Poiseuille, %\cite{Poiseuille:1840}, 
studying the flow of viscous liquids in circular pipes (and capillary tubes in particular), derived the well-known Poiseuille flow profile for a, fully developed, Newtonian fluid. Then, Reynolds %\cite{Reynolds:1883} 
tested experimentally the stability of the Poiseuille profile, finding that it is valid in the case of laminar flow.

Dating from the early twentieth century, the first ones to set up a model for the dynamics of liquid flow into a capillary, were Bell and Cameron \cite{Bell:MLT:1906}, Lucas \cite{Lucas:UZK:1918}, Washburn \cite{Washburn:DCF:1921} and Rideal \cite{Rideal:FLC:1922}. 
The well-known Washburn solution was derived from these works. 
%Subsequently, it was conceived the Bosanquet model, and more recently, another model was the one by Szekely, Neumann and Chang, also called SNC model. 
Washburn solution was found by considering that the motion of a liquid penetrating a capillary is determined by a balance among capillarity, gravitational, and viscous forces under the assumption of a Poiseuille profile as the velocity profile.

For sufficiently large time, Washburn solution describes excellently the dynamics of capillary flow and this is also proved by experimental results. 
Moreover, a further validation of Washburn solution has been given by both simulations of molecular dynamics, see for instance Martic et al. \cite{Martic:MDS:2002,Martic:PDT:2004,Martic:DIP:2005}, and the lattice-Boltzman statistical-mechanical description, mainly used by physicists, see Chibbaro \cite{Chibbaro:CFP:2008}. 
Unfortunately, Washburn solution is defective in describing the initial transient, because the model neglects the inertial effects.
On the contrary, those inertial effects were considered in a model proposed by Bosanquet \cite{Bosanquet:FLC:1923}.
The SNC (Szekely, Neumann and Chang) model, introduced by Szekely et al. \cite{Szekely:RCP:1979},
takes into account the outside flow effects including an apparent mass parameter within the inertial terms.

Recently, many experimental and theoretical studies on liquids flowing
under capillary action have pointed out the limitations of Washburn solution and its validity only as an asymptotic approximation.
For example, liquids flowing in thin tubes were considered 
by Fisher and Lark \cite{Fisher:TIV:1979}. 
Nonuniform cross-sectional capillaries have been studied, for instance by Erickson et al. \cite{Erickson:NSC:2002} and by Young \cite{Young:ACF:2004}.
As far as surface grooves are concerned, we refer to Mann et al. in \cite{Mann:FSL:1995}, Romero and Yost in \cite{Romero:FOC:1996}, 
Rye et al. \cite{Rye:WKS:1996}, and Yost et al. in \cite{Yost:FOC:1997}, 
micro-strips were investigated by Rye et al. \cite{Rye:CFI:1998}. 
On the other hand, several studies have been devoted to capillary rise, dynamics of menisci, wetting and spreading,
see, for instance, the papers by Clanet and Qu\'er\'e \cite{Clanet:OM:2002}, Zhmud et al. \cite{Zhmud:DCR:2000}, Xiao et al. \cite{Xiao:GAC:2006}, Chebbi \cite{Chebbi:DAG:2003}, Fries and Dreyer \cite{Fries:ASC:2008,Fries:DAG:2008} and
the recent books by de Gennes et al. \cite{deGennes:CWP:2004} or by Karniadakis et al. \cite{Karniadakis:MN:2005} and the plenty of references
quoted therein.
Moreover, useful reviews by Dussan \cite{Dussan:SLS:1979}, de Gennes \cite{deGennes:WSD:1985} and Leger and Joanny \cite{Leger:LS:1992}, appeared within the specialized literature.

Our main concern is devoted to report and test the mathematical modelling of entrapped gas action in a horizontal closed-end capillary.
In this context the effect of the entrapped gas on the liquid dynamics
was first studied by Deutsch \cite{Deutsch:PSF:1979} from a theoretical viewpoint and more recently by Pesse et al. \cite{Pesse:ESG:2005} and by Hultmark et al. \cite{Hultmark:IGP:2011} from an experimental one. 

This study is of interest for the non-destructive technique named \lq \lq liquid penetrant testing\rq \rq \ used, for instance, in the production of airplane parts as well as in many industrial applications where the detection of open defects is of crucial interest.

\section{Mathematical modelling}\label{S:model}
\begin{figure}[!hbt]
\centering
\psfrag{O}{$O$}
\psfrag{L}{$L$}
\psfrag{ell}{$\ell$}
\psfrag{R}{$R$}
\psfrag{z}{$z$}
\psfrag{s}[][]{\small $\vartheta$}
\framebox{
\includegraphics[width=.7\textwidth]{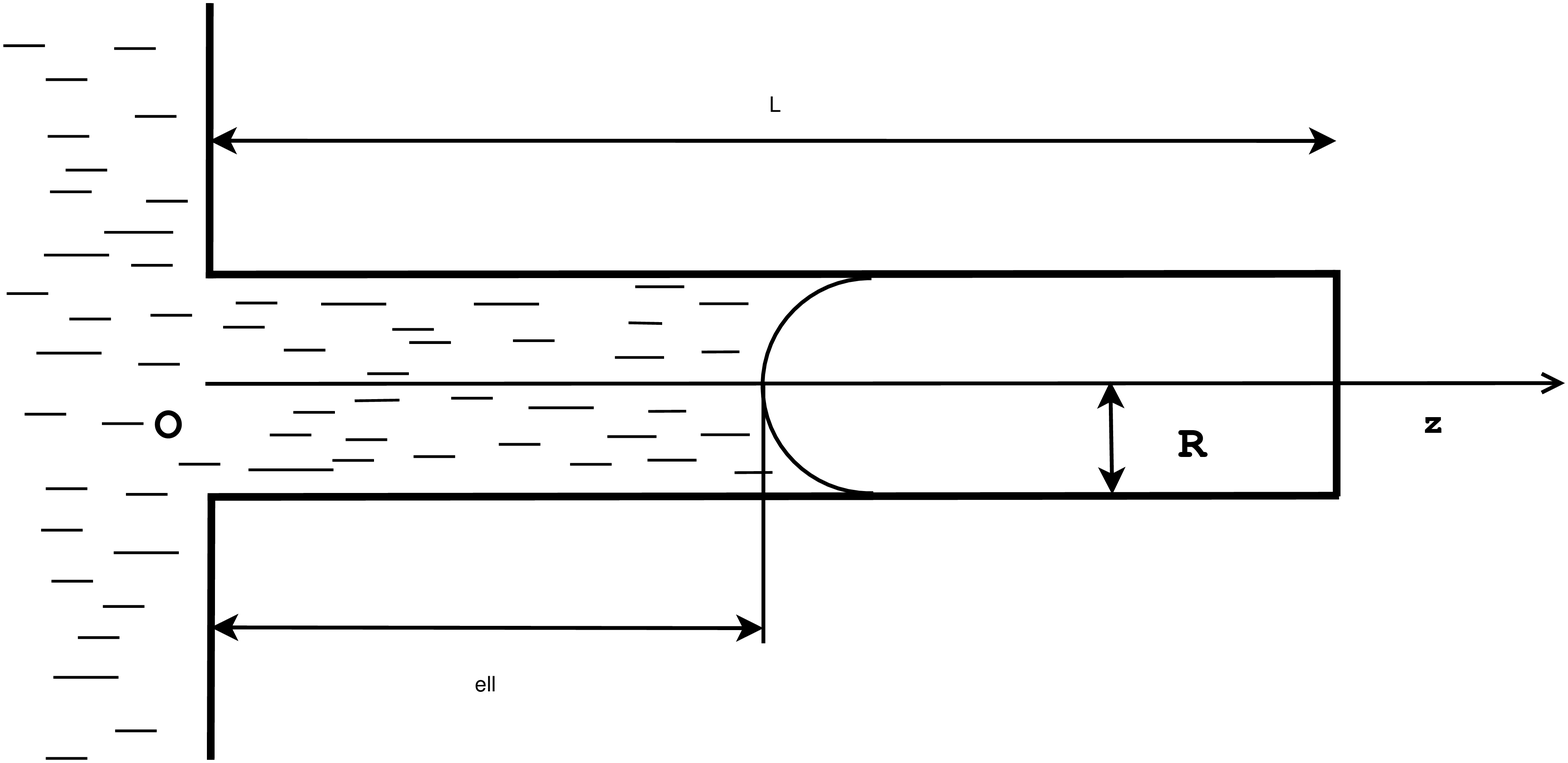}} \\
\framebox{
\includegraphics[width=.7\textwidth]{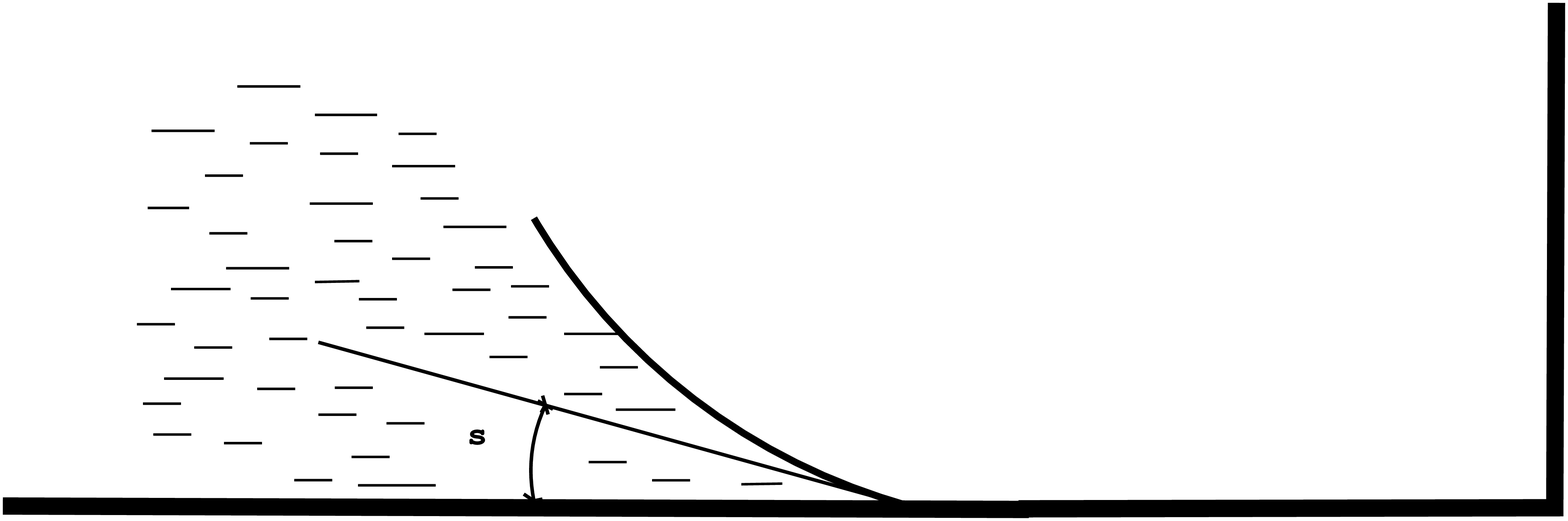}
}
\caption{Top frame: physical setup and notation, the $z$-axis is the horizontal axis. Bottom frame: definition of the contact angle $\vartheta$.}
\label{fig:setup}
\end{figure}

Firstly, let us consider a horizontal cylindrical capillary put in touch with a reservoir, as it is depicted in figure \ref{fig:setup}. 
A formal derivation, for an open-end capillary and two different liquids, from first principles, of a mathematical model describing the capillary dynamics can be found by the interested reader in \cite{Cavaccini:2009:OMN}.
In particular for a horizontal closed-end capillary the Newtonian equation of motion plus natural initial conditions is given by
\begin{eqnarray}\label{eq:model}
\left\{ \begin{array}{l} 
 \rho (\ell + c R) 
	  \displaystyle{\frac{d^2 \ell}{dt^2} + \rho \left(\frac{d\ell}{dt}\right)^2 =	 2 \frac{\gamma \cos \vartheta}{R} 
	  - 8 \frac{\mu \ell}{R^2} \frac{d\ell}{dt}} + \Omega(\ell, L) \ , \\[1.5ex]
\ell(0) =  \displaystyle{\frac{d\ell}{dt}} (0) = 0 \ ,	 
 \end{array}
\right.  
\end{eqnarray}
where $ \ell $ and $L$ are, respectively, the length of the liquid inside the capillary and the capillary length, whereas the remaining parameters assumed constant are:  $\rho$ the density of the liquid, $\gamma$ the surface tension, $\vartheta$ the contact angle, $\mu$ the viscosity, $R$ the capillary radius of the capillary and $ c = O(1) $ the coefficient of apparent mass, introduced by Szekely et al. \cite{Szekely:RCP:1979} in order to get a well-posed problem, see \cite{Fazio:2010:IWP}. 
This coefficient is necessary to make the differential model adequate also for the very steep changes in $\ell$ and its derivative at the very beginning of the dynamics. 
The contact angle $\vartheta$ is defined as the angle defined by the tangent to the meniscus at the point of contact with the capillary wall and the wall itself.
In the case of a closed-end capillary the entrapped gas action
should be taken into account, by specifying to the right
hand side of equation (\ref{eq:model}) the term $\Omega(\ell, L)$ depending on the lengths involved,
whereas for a open capillary $ \Omega(\ell, L) = 0$.

The prescribed initial conditions are discussed at length by Kornev and Neimark \cite{Kornev:SPL:2001}.
As far as the authors knowledge is concerned, no analytical solutions are available for the model (\ref{eq:model}).
On the other hand, recently, asymptotic solutions were derived by Budd and Huang and used to asses the reliability of available numerical schemes, cf. Fazio and Jannelli \cite{Fazio:2010:IWP}.
Moreover, some people assume also, as a further simplification, that the contact angle is constant.
At least in first approximation, such a simplification fits well the condition of a low capillary number and/or $ R/L \ll 1 $, where $L$ is the length of the capillary.
A detailed discussion of the dynamic contact angle simplification is provided elsewhere, see for instance Tokaty \cite{Tokaty:HPF:1994} or Adamson \cite{Adamson:PCS:1997}.

Moreover, in the case of a vertical capillary,
the gravity action should be taken into account, by adding to the left
hand side of equation (\ref{eq:model}) the term
\begin{eqnarray}
\pm (\rho \ell) g \ , 
\end{eqnarray}
where the plus or minus sign has to be used when the liquid reservoir
is over or below the cavity, respectively.

%We assume that the only physical entities active on bulk of liquid are: the surface tension, the viscosity, the pressure of the entrapped gas and the atmospheric pressure.
For the validity of our one-dimensional analysis we must assume that the fluid has a quasi-steady Poiseuille velocity profile. 
This consists in assuming that the fluid motion is a laminar flow, i.e. the liquid is considered to be moving in circular concentric circles with a parabolic velocity profile null at the wall, the so-called no-slip boundary condition.

\subsection{Entrapped gas modelling}
Two different ways to model the entrapped gas action are available in literature:
the first, given by the following formula  
\begin{eqnarray}\label{eq:pe1}
\Omega(\ell, L) = p_a - p_a \frac{L}{L - \ell} \ , 
\end{eqnarray}
is due to Deutsch \cite{Deutsch:PSF:1979}, here $ p_a $ is the atmospheric pressure;
the second, according to Zhmud et al. \cite{Zhmud:DCR:2000}, Chibbaro \cite{Chibbaro:CFP:2008} or Hultmark et al. \cite{Hultmark:IGP:2011}, takes into account only the 
viscous drag produced by the entrapped gas as follows
\begin{eqnarray}\label{eq:pe2}
\Omega(\ell, L) = -\frac{8 \mu_g (L-\ell)}{R^2} \frac{d\ell}{dt}- \frac{d}{dt}\left[\rho_g (L-\ell)\frac{d\ell}{dt}\right] \ , 
\end{eqnarray}
where $ \mu_g $ and $\rho_g$ are the viscosity and density of the entrapped gas.
Let us remark that those authors usually take $\mu_g \ll \mu$.

By getting closer to the physics of the phenomenon, the gas action occurring inside the capillary can be seen as an adiabatic compression. 
This implies the assumption of no heat exchange during the liquid penetration. Under this assumption, we can use the adiabatic equation of a gas expressed by
\[p\; V^\lambda=constant \ ,\]
where $p$ is the pressure, $V$ is the gas volume, and $\lambda>0$ is the well known gas constant ($\lambda =1.4$ for bi-atomic gases) represented by the ratio between the  specific heat at constant pressure and the specific heat at constant volume. 
In order to deduce the value for the constant, we can assume that $p_a$ is the pressure at time $t=0$ when the volume occupied by the gas is equal to $A \cdot L$, where $ A = \pi R^2 $ denotes the cross sectional area of the capillary. 
As a result, we have that the constant is equal to $p_a A^\lambda L^\lambda$ and the adiabatic equation becomes 
\[p(t)(L-\ell)^\lambda A^\lambda=p_a L^\lambda A^\lambda \ .\]
The entrapped gas action can be taken into account by 
adding to the right-hand side of equation
(\ref{eq:model}) the pressure that this gas applies to the 
penetrant liquid meniscus, that is 
\begin{eqnarray}\label{eq:pe3}
\Omega(\ell, L) = p_a - p_a \left(\frac{L}{L - \ell}\right)^{\lambda} \ .
\end{eqnarray}
It is evident that the model by Deutsch is a specific case of the adiabatic model obtained by fixing $\lambda=1$, that is about the value for dry air. 

Let us consider here equation (\ref{eq:pe3}), and remark that the entrapped gas pressure
%\begin{equation}
\[p_e = p_a \left(\frac{L}{L - \ell}\right)^{\lambda} \]
%\end{equation}
verifies the initial condition $ p_e(\ell = 0) = p_a $, 
the asymptotic condition
%\begin{equation}
\[\lim_{L \rightarrow \infty } p_e = p_a \ ,\]
%\end{equation}
and the limit condition
%\begin{equation}
\[\lim_{\ell \rightarrow L} p_e = + \infty \]
%\end{equation}
corresponding to the common intuition that 
no action is expected if the capillary is not closed and that 
the internal pressure will increase if we let $ \ell $ increase for a closed capillary.

In the study of capillary dynamics, it is of primary interest to work out the stationary level reached by the fluid inside the capillary.
This level corresponds to the steady-state solution, obtained from (\ref{eq:model:ad}) by setting to zero velocity and acceleration.
The inflow will cease at $\ell_{max}$ with
\begin{eqnarray}
     \frac{\ell_{\max}}{L}=\frac{2\gamma \cos\vartheta}{R\; p_a+2\gamma \cos\vartheta} \ .
\label{eq:zsul}
\end{eqnarray}
In particular, if we consider the case $ \lambda = 1 $, according to Deutsch \cite{Deutsch:PSF:1979}, who used $L=1$, $ R = 10^{-4}$m %($\approx 1\times10^6$ dyncs/cm$^2$)
and an air-water interface with $ \gamma = 71.8 \; $ $10^3$N/m, $ \vartheta = 0^\circ $, and $p_a$ equal to one atmosphere ($\approx 10^5$ N/m$^2$), equation (\ref{eq:zsul}) shows that the flow will cease at
\begin{eqnarray}\label{eq:15}
    \frac{\ell_{max}}{L}\approx1.5\% \ ,
\end{eqnarray}
and there will be about 98.5\% of the capillary depth left to be filled.

\begin{figure}[!hbt]
\centering
\psfrag{x}{$t$}
\psfrag{y}[br][br][1][-90]{$\displaystyle{\frac{\ell(t)}{L}}$\quad}
\psfrag{a1}{$\lambda=1$}
\psfrag{a2}{$\lambda=1.1$}
\psfrag{a3}{$\lambda=1.2$}
\psfrag{a4}{$\lambda=1.3$}
\psfrag{a5}{$\lambda=1.4$}
\includegraphics[width=.7\textwidth,height=10cm]{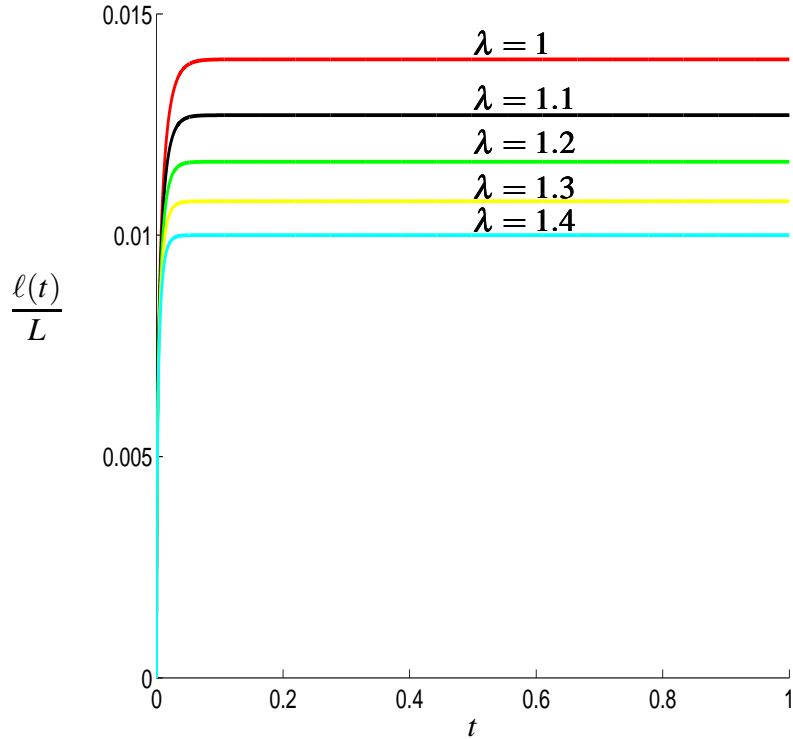} 
\caption{Water simulation in a close-end capillary. Equation (\ref{eq:pe3}) is used with different values of $ \lambda$.}
\label{fig:water}
\end{figure}

\section{Numerical results}
Of course, we have also performed several numerical tests, involving the parameters 
characterizing different real liquids as well as the entrapped gas pressure term.
We used the liquid parameters listed in \cite{Kornev:SPL:2001} and reported, for the reader convenience, in table~\ref{rfazio:table2}.
\begin{table}[!hbt]
\centering
		\begin{tabular}{lr@{}lclr}
	  \hline
	   & \multicolumn{2}{c}%
	     {}        &             & Surface & Contact \\
	   & \multicolumn{2}{c}%
	   {Viscosity} $ \mu $ & Density $ \rho $ & tension $ \gamma $ & angle $ \vartheta $ \\
	  Liquid & \multicolumn{2}{c}%
	  {(mPa$\cdot$s)}   & (Kg/m$^3$)  & (mN/m)     &  \\
	  \hline		
	  Silicon oil & 500&    & 980 & 21.1 & $ 0^\circ $  \\
	  Ethanol     &   1&.17 & 780 & 21.6 & $ 0^\circ $   \\
	  Ether       &   0&.3  & 710 & 16.6 & $ 0^\circ $   \\
	  Mixture     &   0&.77 & 955 & 57   &$ 10^\circ $ \\
	  Water       &   1&    & 998 & 71.8 & $ 0^\circ $  \\
	  \hline
		\end{tabular}
		\caption{Parameters of different liquids according to the recent survey by Kornev and Neimark \cite{Kornev:SPL:2001}.}
    \label{rfazio:table2}
\end{table}
The interested reader can find a similar table related to the so called PDMS silicone olis in Fazio et al. \cite{Fazio:2012:ESI}.

As a simple test case, figure \ref{fig:water} shows the numerical solution, for different values of $\lambda$, corresponding to a closed-end capillary and water, obtained for the same values used by Deutsch \cite{Deutsch:PSF:1979} and reported in the end of the previous section.
%$p_a=101325 Pa$, $\rho=998 Kg/m^3$, $\mu=10^{-3} Pa\cdot s$, $\gamma=71.8\cdot 10^{-3} N/m$, $\vartheta=0.0175 rad$, $R=10^{-4} m$, $ \lambda = 1$ and $L=1 m$.

\begin{figure}[!hbt]
\centering
\psfrag{x}{$t$}
\psfrag{y}[br][br][1][-90]{$\displaystyle{\frac{d\ell}{dt}(t)}$\quad}
\includegraphics[width=.75\textwidth]{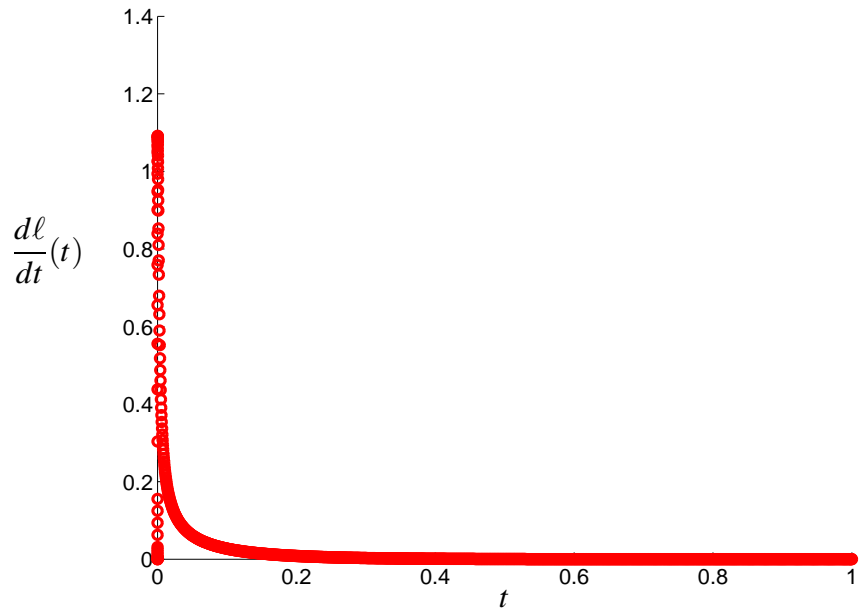} 
%\put(-185,5){\Large $t$}
%\put(-375,125){\Large $\displaystyle{\frac{d\ell}{dt}(t)}$}
\caption{Transitory of $ \frac{d\ell}{dt}(t)$ for $ \lambda = 1 $.}
\label{fig:transitory}
\end{figure}

%Here, we used the (\ref{eq:pe3}) relation.
Figure \ref{fig:transitory} shows a sample behaviour for the inclusion velocity (first derivative of the front length). 
  
\begin{figure}[!hbt]
\centering
\psfrag{y}[br][br][1][-90]{$\displaystyle{\frac{\ell(t)}{L}}$}
\psfrag{x}{$t$}
\includegraphics[width=.75\textwidth]{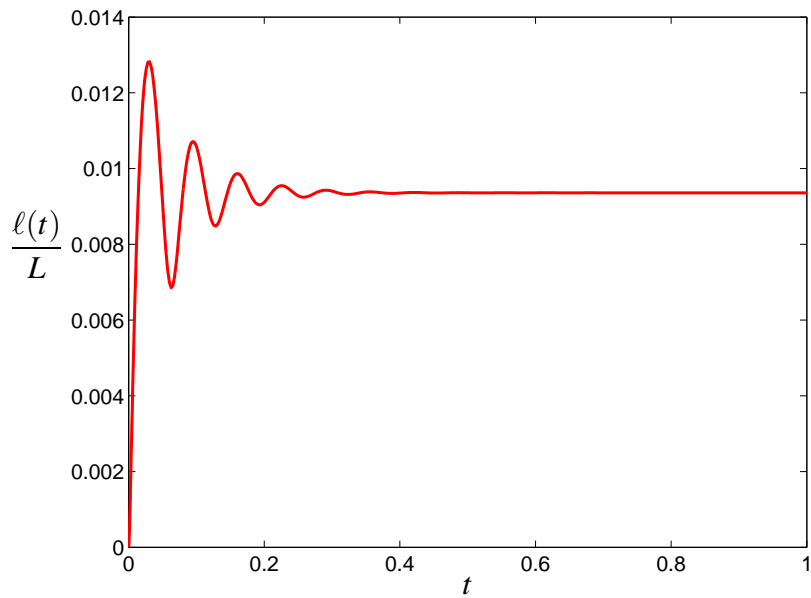} 
%\put(-185,0){\Large $t$}
%\put(-385,125){\Large $\displaystyle{\frac{\ell(t)}{L}}$}
\caption{Typical oscillations for a meniscus inside a close-end cylindrical capillary.}
\label{fig:oscill}
\end{figure}

The case $ \lambda = 1 $ was already treated in \cite{Cavaccini:2009:OMN}.
We remark that our obtained result, i.e. $\ell_{\max} \approx 1.4\% $, is thinly different from the one of Deutsch in equation (\ref{eq:15}). 
Depending on the values used for the geometrical parameteres $L$ and $R$, by keeping fixed all the other parameters affecting the dynamics, we have also observed the occurrence of oscillatory damped solutions. 
The possibility to obtain this kind of solution was pointed out also by Zhmud et al. \cite{Zhmud:DCR:2000}. 
In particular, it was numerically observed that such solutions oscillate just around the stationary level reached by the liquid interface and that, keeping fixed all the parameters, it always exists a threshold for the radius fixed the total length and viceversa, so that, the dinamics passes from an oscillatory under-damped behaviour to an over-damped one, or the other way round.
%A sample of this behaviour for $Ca =0.002$, $Re =50$, $P = 6.5\cdot 10^3$, $\lambda=1$, $R =2\cdot 10^{-4}$, and $L=0.1$ is shown in figure \ref{fig:oscill}. 
A sample of this behavior for $\lambda=1$ and for water inside a cylindrical capillary, with $L=0.1$ m and $R=1.5\cdot 10^{-4}$ m, is shown in figure \ref{fig:oscill}. 

%From a qualitative viewpoint, the term responsible for the presence, of the damped oscillations observed in figure \ref{fig:oscill} rather than the non oscillatory behavior in figure \ref{fig:water}, is the viscous term, i.e. the one in (\ref{eq:model}) containing the constant $\mu$. %and proportional to the square of the capillary radius $R$. 
%Indeed, having fixed a finite capillary length $L$, from numerical experiments it can be deduced the existence of a threshold on the capillary radius, so that, for higher values there is a non oscillatory behavior instead of the oscillatory one obtained for lower values.
Anyway the analytical determination of this transition threshold will be studied in deep in the next section.

The numerical results reported in this section were obtained by the \texttt{ODE45} solver of the MATLAB ODE suite developed by Shampine and  Reichelt \cite{Shampine:MOS:1997}.

\section{Analytical study of the capillary rise}
We have found, numerically, the existence of a threshold for the occurrence of an oscillatory behaviour. 
Moreover, the recent existing literature  \cite{Fries:DAG:2008} on the derivation of an analytical solution of capillary rise unfortunately deals with a model subjected to gravity but not taking into account the presence of an entrapped gas. 
For this reason, it is our opinion that an analytical study for the present problem taking into account the presence of an entrapped gas migth be of great interest. 
This topic was already partially faced in \cite{Fazio:2009:EGA}.
By setting the apparent mass coefficient $c=1$, the complete model, including the action of the entrapped gas (\ref{eq:pe3}), with $\lambda=1$, can be rewritten as follows
\begin{equation}\label{eq:closed-end-expanded}
\left(\displaystyle{\frac{d\ell}{dt}}\right)^2+\ell\displaystyle{\frac{d^2\ell}{dt^2}}+R\displaystyle{\frac{d^2\ell}{dt^2}}=\frac{a}{R}-2b\frac{\ell}{R^2}\displaystyle{\frac{d\ell}{dt}}-\frac{P\ell}{L-\ell} \ ,
\end{equation}
with the initial conditions $ \ell(0)=\displaystyle{\frac{d\ell}{dt}}(0)=0$, the new parameters 
\[
a=2\frac{\gamma cos(\theta)}{\rho}\qquad b=4\frac{\mu}{\rho}\qquad P=\frac{P_a}{\rho} \ .
\]

\subsection{Non-oscillatory regime}
\begin{figure}[!htb]
	\centering
	\psfrag{x}{$s$}
	\psfrag{Ep}[br][br][1][-90]{$E_p$\quad}
	\psfrag{R=2e-3}{\tiny $R=2 \cdot 10^{-3}$}
	\psfrag{R=2e-4}{\tiny $R=2 \cdot 10^{-4}$}
	\psfrag{R=2e-5}{\tiny $R=2 \cdot 10^{-5}$}
		\includegraphics[width=.65\textwidth]{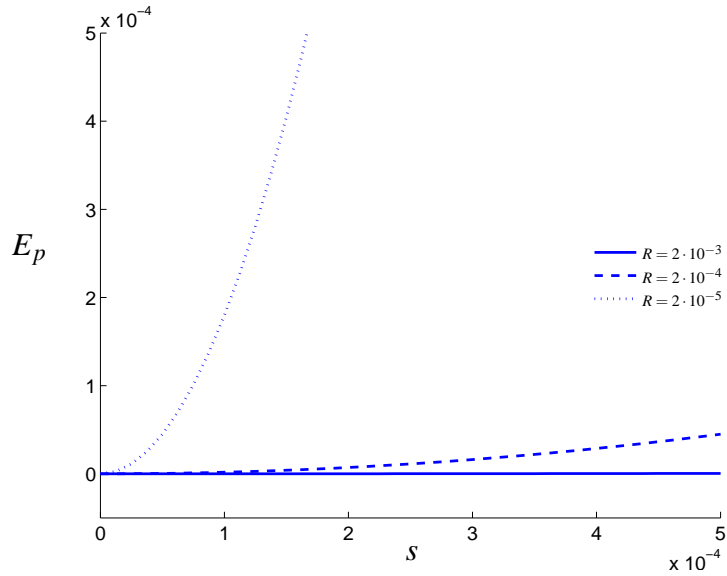}
%\put(-175,5){\Large $s$}
%\put(-359,135){\Large $E_p$}
	\caption{Error function $E_p$ versus $s$.}
	\label{fig:errpar}
\end{figure}

For the sake of calculation and numerical simplification, it is worth rescaling the problem by introducing the generic scaling group
\begin{eqnarray} 
\ell=R^\alpha v\qquad t=R^\beta s \ ,\label{eq:scaling}	
\end{eqnarray}
with $\alpha$ and $\beta$ to be determined by replacing the original variables in (\ref{eq:closed-end-expanded}).
As a matter of fact, we get 
\[ R^{2\alpha-2\beta}\left(\left(\frac{dv}{ds}\right)^2+v\frac{d^2v}{ds^2}\right)+R^{\alpha-2\beta+1}\frac{d^2v}{ds^2}=R^{-1}a-2bR^{2\alpha-\beta-2}v\frac{dv}{ds}-PR^\alpha v/(L-R^\alpha v) \ .
\]
As we have the forcing term that is $R^{-1}a$, we are somehow forced to make this exponent appear also in the last term on the left hand side and in the second one on the right hand side of the equation. 
This is just obtained by solving the linear system
\begin{equation}
\left\{\begin{array}{ll}
2\alpha-\beta-2=-1\\
\alpha-2\beta+1=-1 \nonumber
\end{array}\right.
\end{equation}
that, once solved, provides the solution 
\[ 
\alpha =4/3 \qquad \beta=5/3 \ , 
\]
so that (\ref{eq:scaling}) becomes
\[
\ell=R^{4/3}v\qquad t=R^{5/3}s \ .
\]
In such a way we get the 'scaled form' for (\ref{eq:closed-end-expanded})
\begin{equation}\label{eq:closed-end-expanded-scaled}
R^{-2/3}\left[\displaystyle{\left(\frac{dv}{ds}\right)^2+v\frac{d^2v}{ds^2}}\right]+R^{-1}\displaystyle{\frac{d^2v}{ds^2}}=R^{-1}a-R^{-1}2bv\displaystyle{\frac{dv}{ds}}-\displaystyle{\frac{PR^{4/3}v}{L-R^{4/3}v}} \ ,
\end{equation}
subject to the initial conditions $ v(0)=\displaystyle{\frac{dv}{ds}}(0)=0$.
At this point, on the left hand side of (\ref{eq:closed-end-expanded-scaled}) we have the first term that is proportional to $R^{-2/3}$, whereas the one on the right hand side relative to the action of the entrapped gas, as long as the penetration is far from  filling the capillary, is proportional to $R^{4/3}$, and the remaining three terms are proportional to $R^{-1}$. 
Making tend $R$ to zero it is clear that the dominant terms are just the ones proportional to $R^{-1}$. 
Hence, by ignoring  the terms not proportional to $R^{-1}$, the leading equation for (\ref{eq:closed-end-expanded-scaled}) results 
$$\frac{d^2v}{ds^2}=a-2bv\frac{dv}{ds}$$ that integrated once, due the given initial conditions, becomes
\begin{eqnarray}\label{eq:leading}
\frac{dv}{ds}=as-bv^2 \ .
\end{eqnarray}
Equation (\ref{eq:leading}) belongs to the class of Riccati equation and is well posed.
Furthermore, as for very small $s$, being zero the function $v$ in the origin, the second term on the right hand side of (\ref{eq:leading}) can be ignored so as in the very beginning of its evolution, the solution can be approximated as follows
\begin{eqnarray}
v\approx as^2/2 \quad \mbox{for}\quad s\ll1.% \quad \mbox{valid }s<1 \mbox{i.e.} t<R^{5/3},
\end{eqnarray}
By indicating the numerical solution as $v_n$, obtained for the same parameters of water already used in figure \ref{fig:oscill}, and defining the function $E_p(s)=\left|v_n-a\frac{s^2}{2}\right|$ as well, in figure \ref{fig:errpar} it can be appreciated how in the very beginning of its dynamics the solution can be assimilated to a parabola within a very high degree of precision. 
Moreover, it turns out also that the more is the capillary radius, the longer is the time interval of validity of the above deduction. 

\begin{figure}[htbp]
	\centering
	\psfrag{x}{$s$}
	\psfrag{Ew}[br][br][1][-90]{$E_w$\quad}
	\psfrag{R=2e-3}{\tiny $R=2 \cdot 10^{-3}$}
	\psfrag{R=2e-4}{\tiny $R=2 \cdot 10^{-4}$}
	\psfrag{R=2e-5}{\tiny $R=2 \cdot 10^{-5}$}
		\includegraphics[width=.65\textwidth]{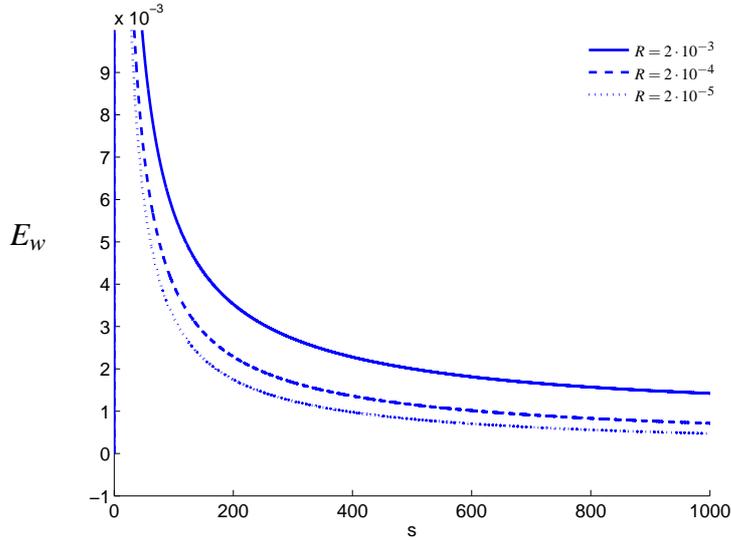}
	\caption{Error function $E_w$ versus $s$.}
	\label{fig:errwash}
\end{figure}

On the contrary, as $v$ is increasing, after some time it becomes so high that we are allowed to ignore the left hand side of (\ref{eq:leading}) so as the solution can be given by
\begin{eqnarray}
v\approx\sqrt{\frac{as}{b}}\quad \mbox{for}\quad s\gg1  \mbox{\qquad Washburn solution}.
\end{eqnarray}
Analogously to what done above, by defining the function $E_w(s)=\left|v_n-\sqrt{\frac{as}{b}}\right|$, in figure \ref{fig:errwash} it can be observed how after a given time where, as seen above there is a parabolic trend, unless the solution has reached the steady state (this happens for $R<10^{-4}$), there is a time interval widening as the capillary radius lowers where Washburn solution represents an excellent approximation. 

\subsection{Oscillatory regime}
\begin{figure}[!hbt]
\centering
\psfrag{Lt}[br][br][1][-90]{${L_t}$}
\psfrag{rt}{$R_t$}
\psfrag{A}{{non-oscillatory}}
\psfrag{B}{{oscillatory}}
\includegraphics[width=.75\textwidth]{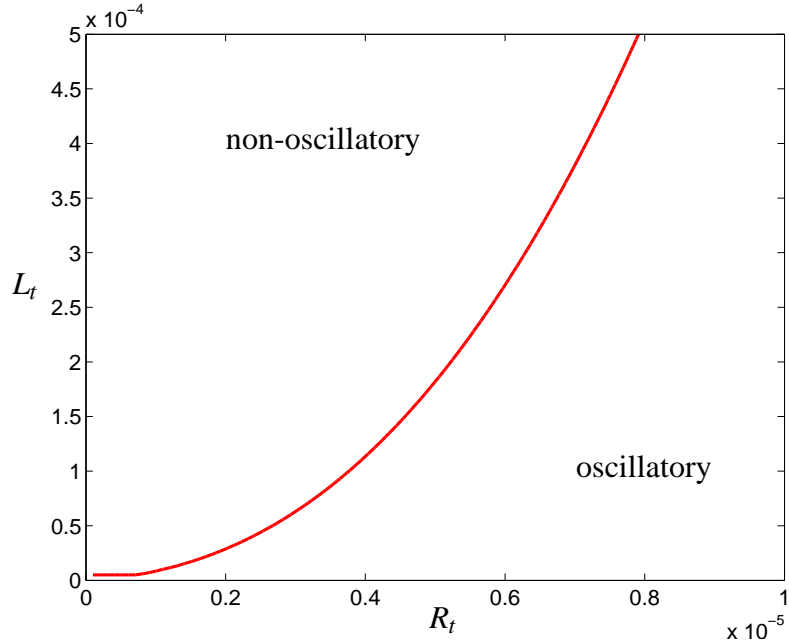} 
%\put(-185,0){\Large $t$}
%\put(-385,125){\Large $\displaystyle{\frac{\ell(t)}{L}}$}
\caption{Oscillatory and non-oscillatory regions for water.}
\label{fig:o-nono}
\end{figure}

By recalling (\ref{eq:zsul}) in term of the new parameters $a$ and $b$, 
we are able to linearize the original equation around the steady state level, by replacing $\ell(t)=\ell_{max}+\epsilon(t)$ in (\ref{eq:closed-end-expanded}), where $\left|\epsilon(t)\right|\ll \ell_{max} $ for all $t$ is just a small perturbation.
By ignoring the terms in $\epsilon$ with order higher than one it results 
\begin{eqnarray}
(\ell_{max}+R)\ddot{\epsilon}+\frac{2b\ell_{max}}{R^2}\dot{\epsilon}+\frac{PL}{(L-\ell_{max})^2}\epsilon=0.
\label{eq:linear}
\end{eqnarray}
By using this linearized second order equation, we will understand how to estimate analytically the threshold value for the capillary radius or the capillary length about which some oscillations starts showing up around the steady state level $\ell_{max}$, or alternatively the motion tends, without any oscillation, asymptotically to $\ell_{max}$. 
By imposing the zero discriminant condition on the characteristic polynomial associated to (\ref{eq:linear}), an equation in two variables, namely, the capillary radius and length can be easily worked out, provided that we keep on maintaining constant all the remaining parameters.
Such a condition is just represented by the following equation
\[
\frac{b^2\ell_{max}^2}{R^4}=\frac{P L}{(L-\ell_{max})^2}(\ell_{max}+R) \ ,
\]
which after some algebraic manipulations, and by recalling the functional dependence of $\ell_{max}$ in (\ref{eq:zsul}), can be rewritten as the polynomial of seventh degree in $R$ and third degree in $L$
\begin{eqnarray}\label{eq:transequation}
b^2 a^2 L^3 P-R^2\left(a+RP\right)^3\left(a L+R^2 P + a R\right) = 0 \ ,
\end{eqnarray}
By regarding (\ref{eq:transequation}) as a two variables function $R$ and $L$, its zero level defines a curve on the $(R_t,L_t)$ plane reported in figure (\ref{fig:o-nono}). 

\begin{figure}[!hbt]
\centering
\psfrag{Lt}[br][br][1][-90]{${L_t}$}
\psfrag{rt}{$R_t$}
\psfrag{A}{{non-oscillatory}}
\psfrag{B}{{oscillatory}}
\includegraphics[width=.75\textwidth]{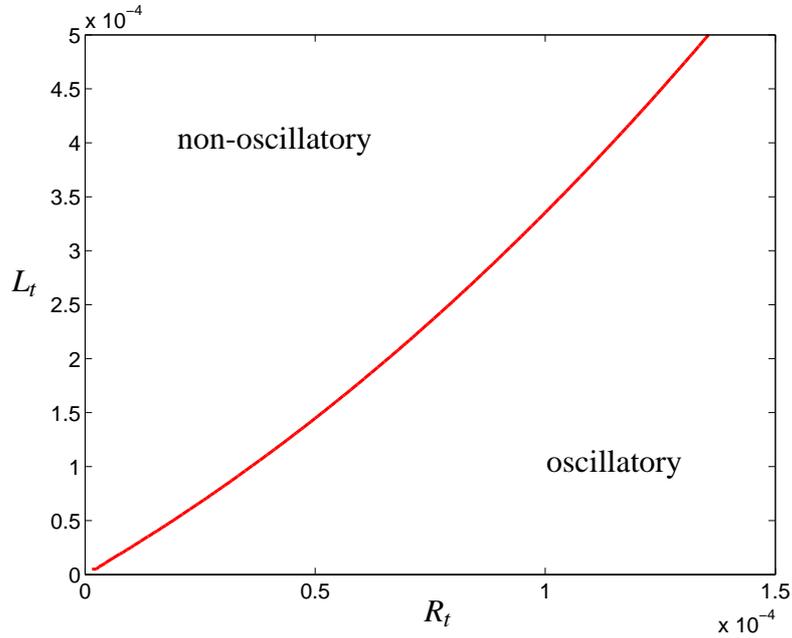} 
%\put(-185,0){\Large $t$}
%\put(-385,125){\Large $\displaystyle{\frac{\ell(t)}{L}}$}
\caption{Oscillatory and non-oscillatory regions for silicone oil.}
\label{fig:o-nono-sil}
\end{figure}

Furthermore, if we keep constant also the length or the radius, by means of any root finder, we can compute the corresponding radius and length threshold respectively. 
By changing the type of penetrant liquid from water to silicone oil having the physical features reported in table (\ref{rfazio:table2}), it can be worked out the picture in figure (\ref{fig:o-nono-sil})
 
\begin{figure}[htbp]
	\centering
	\psfrag{dell}[bb][bb][1][-90]{$\displaystyle{\frac{d\ell}{dt}}$ \quad}
	\psfrag{ell}[tt][tt][1][0]{$\ell$\quad}
	\psfrag{R=0.0021}{\tiny $R=2.1 \cdot 10^{-3}$}
  \psfrag{R=0.0023}{\tiny $R=2.3 \cdot 10^{-3}$}
  \psfrag{R=0.0025}{\tiny $R=2.5 \cdot 10^{-3}$}
  \psfrag{R=0.0027}{\tiny $R=2.7 \cdot 10^{-3}$}
  \psfrag{R=0.0029}{\tiny $R=2.9 \cdot 10^{-3}$}
	\psfrag{linf}{\tiny{$\ell_\infty$}}
	\includegraphics[width=.65\textwidth]{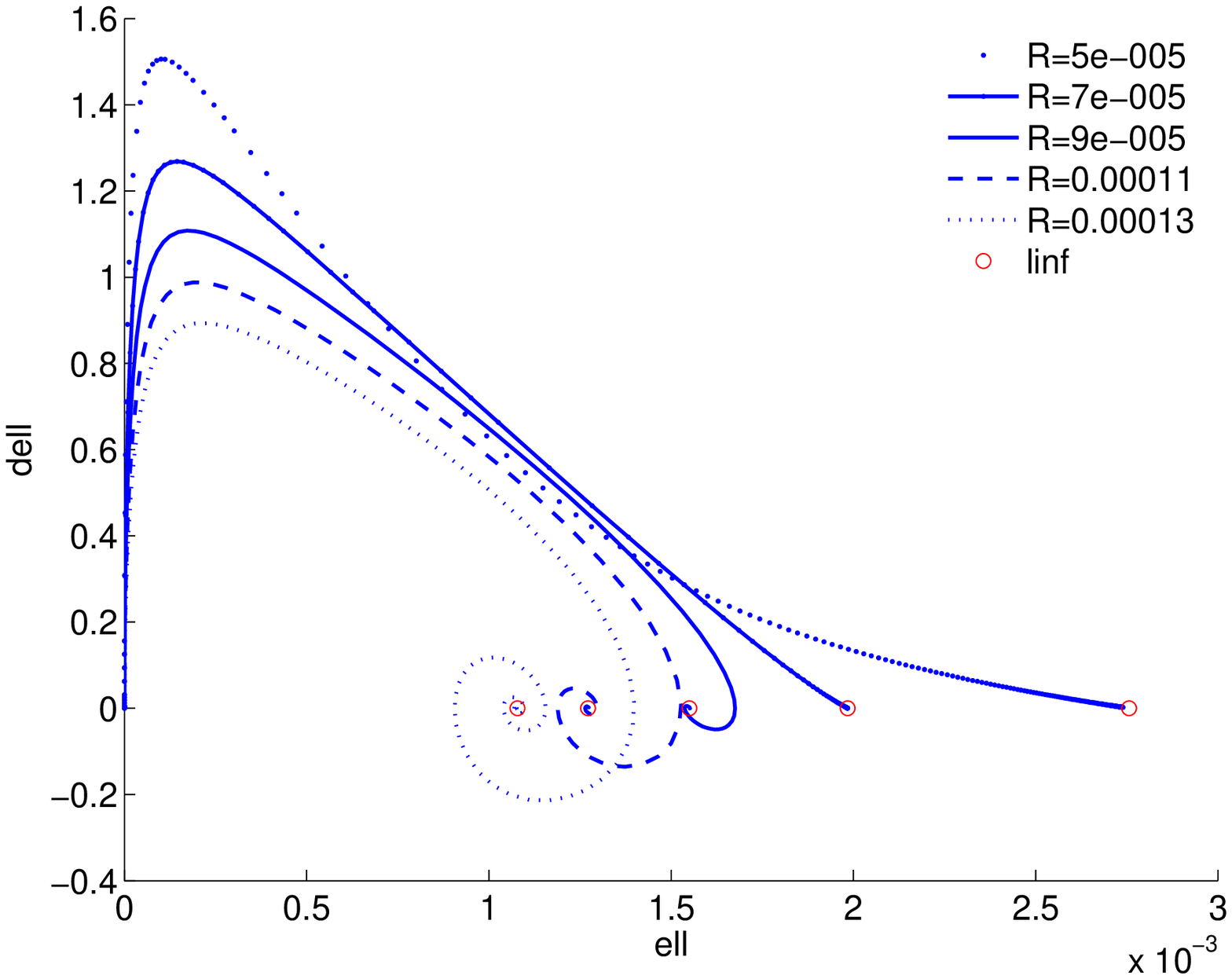}
	\psfrag{dell}[bb][bb][1][-90]{$\displaystyle{\frac{dv}{ds}}$\quad} \\
	\psfrag{ell}[tt][tt][1][0]{$v$}
	\includegraphics[width=.65\textwidth]{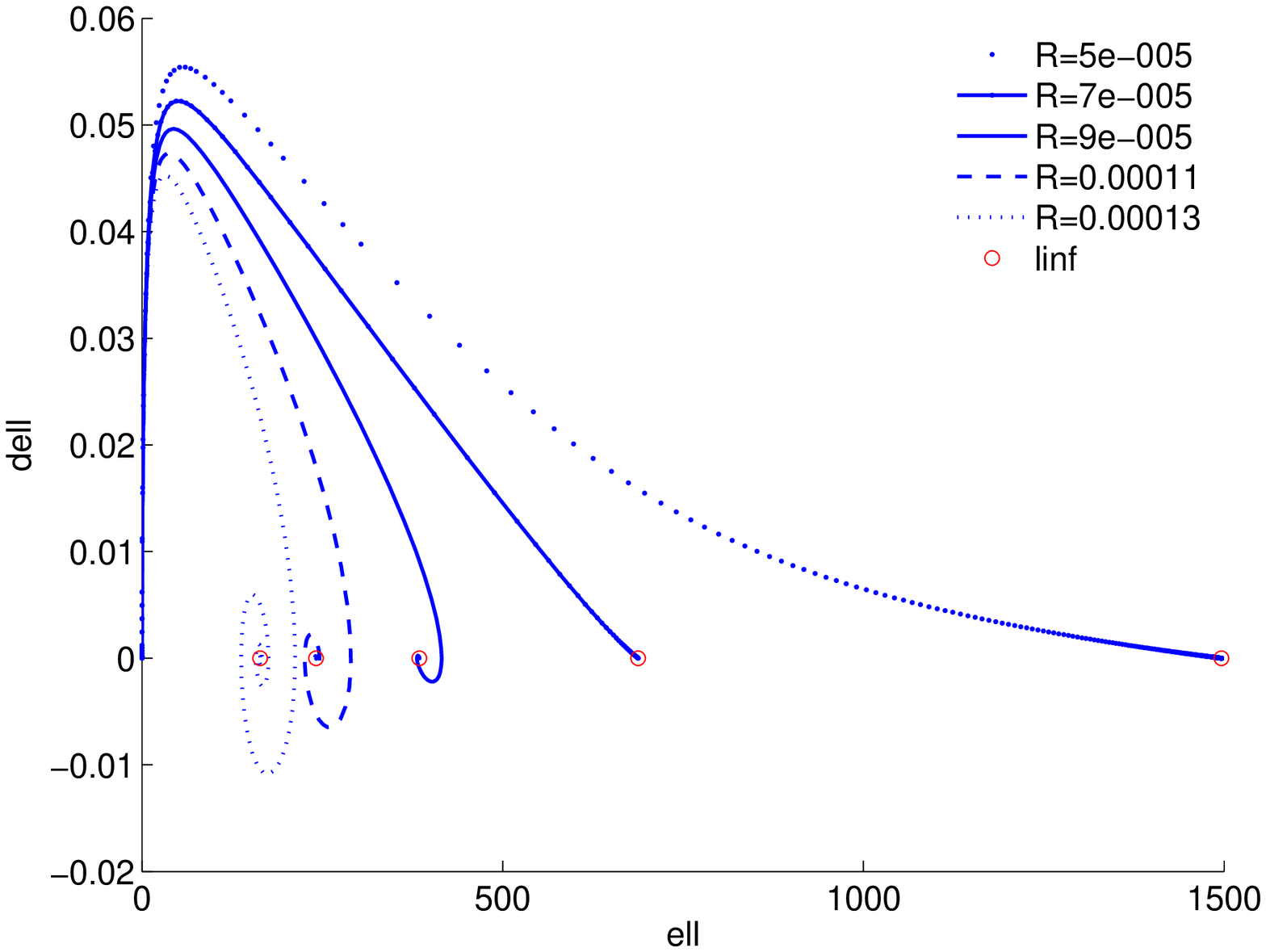}
	\caption{Phase plane transition from under-damped to over-damped regime. On the top frame: for the unscaled model (\ref{eq:closed-end-expanded}) and for the scaled model (\ref{eq:closed-end-expanded-scaled}) on the bottom frame.}
	\label{fig:phaseplane}
\end{figure}

This can be also pictorially observed by looking at figure \ref{fig:phaseplane} where the trajectories for a set of radius values, including the ones just said, are shown: on the top frame the unscaled function $\ell(t)$ and on the bottom frame the scaled function $v(s)$, respectively.
% It means that relative error $E=(R_t-R_n)/R_t=1.057$. 
Moreover, it can be also seen how operating with the scaled variables $v(s)$ the transition can be better appreciated as the curves do not cross each other.
  %, or in percentage terms the calculated value is 5.7\% higher than the numerical one.
 
Just for sake of completeness, the same calculation has been carried out for the oscillatory example already reported in figure \ref{fig:oscill}. 
In this case the corresponding parameters are $a=1.44 \cdot 10^{-4}$, $b=4.01 \cdot 10^{-6}$, $P=101.528$ so that for them it comes out a theoretical threshold $R_t=7.275 \cdot 10^{-5}$ in a very  agreement with the numerical one. %deduced by starting from a value, expressed with two decimal figures, high enough for oscillations and through its successive reductions it has been chosen its first value where the derivative variations are under $10^{-6}$.

%The difference between the theoretical and numerical values can be explained by means of the truncation error that we introduce when we linearize the differential model: this error is more remarkable for higher values of surface tension and viscosity, such as in the first academic simulation, than it is actually for values of the parameters much closer to values of real physical situations.
Finally, it is worth adding that the same results on the derivation of the analytical threshold have also been obtained by means of a stability analysis through the study of the Jacobian of the vector field around the critical point represented by the steady state level on the phase plane $(\ell_{max},0)$. In this case it has been confirmed that $(\ell_{max},0)$ is a stable point but just for the same threshold value found above it changes its nature from a simple node to a spiral node.

\section{Conclusions}
%\subsection{Adimensional model}
Before drawing the conclusions for our study it is worth mentioning how our model (\ref{eq:model})-(\ref{eq:pe3}) can be made adimensional. 
There are two main reasons that make an adimensional model important. 
As a first reason, the number of independent paramaters is reduced. 
The so-called pi-theorem tells us that for each characteristic dimension a reduction of one variable is possible. 
A second, and possibly more important, reason is that we get a dimensionaless parameters which characterize
the problem by the value they have. 
Each parameter represent a particular aspect of the problem. 
If its value is around unity, then that particular aspect is important.
But, on the contrary, if its value is extreme, i.e. close to zero, or very large then that particular aspect can be treated in an asymptotic sense.

As far as our model (\ref{eq:model})-(\ref{eq:pe3}) is concerned, by setting $c=1$, introducing the new dimensionless variables $x = \ell/L$, $\tau = t U/L$, and the parameters $\delta = R/L$, the capillary number $ Ca = \mu U /\gamma $, the Reynolds number $Re = 2R U / \mu$ and $P = p_a/(\rho U^2)$, where $U$ is the average velocity, we get the adimensional model
\begin{eqnarray}\label{eq:model:ad}
\left\{ \begin{array}{l} 
 (x + \delta) 
	  \displaystyle{\frac{d^2 x}{d\tau^2} + \left(\frac{dx}{d\tau}\right)^2 =	 4 \frac{\cos \vartheta}{Ca\; Re} 
	  - 16 \frac{x}{\delta \; Re} \frac{dx}{d\tau}} + P - \frac{P}{(1-x)^\lambda} \ , \\[1.5ex]
x(0) =  \displaystyle{\frac{dx}{d\tau}} (0) = 0 \ .	 
 \end{array}
\right.  
\end{eqnarray}
%Moreover, we require that $ Re = 2R U / \mu \ll 1 $, $ Bo = \rho g R^2/\gamma \ll 1 $, $ Ca = \mu U /\gamma \ll 1 $, and $ We = 2 \rho R U^2/\gamma \ll 1 $.
It is worth briefly explaining the meaning of these parameters:
\begin{itemize}
\item A capillary number $Ca$ smaller than one means that the surface tension is predominant over the viscous effect for slow fluid motion.
\item A Reynolds number $Re$ lower than 2300 indicates the physical condition for laminar flow.
\end{itemize}
Moreover, it would be also possible to introduce other two adimensional fluidynamics paremeters: the Bond and the Weber numbers.
The lower is the Bond number $Bo= \rho g R^2/\gamma$, the less is the deformation induced by the gravitational acceleration on the spherical shape of the meniscus interface due to the surface tension, in a horizontal cylinder.
A low Weber number $We = Ca \; Re$ means that the inertia of fluid is negligible with respect to its surface tension.

Afther a deeper study we have found that our analysis is better undertaken within the dimensional model. For instance, the scaling analysis presented in the previous section cannot be carried on for the adimensional model because a value of $R$ appearing on the first two terms on the right-hand side of our model is replaced by the Reynolds number. 
Moreover, all dimensional parameters of different penetrant liquids can be taken from tables available in the literature, as for instance in the recent survey by Kornev and Neimark \cite{Kornev:SPL:2001} or in Fazio et al.
\cite{Fazio:2012:ESI}, whereas in order to correlate a real liquid to the adimensional model we need to find out a characteristic mean velocity U and this is not available within the relevant literature.

The dynamical behaviour of a liquid inside a capillary entrapping a gas represents a very challenging phenomenon, from a modelling, an analytical, and a numerical viewpoint. 
In this paper it has been proved how the specific features of the liquid as well as of the capillary dimensions and the entrapped gas affect the dynamics of capillary rise in terms of time, depth, velocity of penetration, and type of dynamical regime (oscillatory or non-oscillatory). 
Moreover, it has been provided a new tool to determine just the threshold for the transition from one dynamical regime to the other in reaching the steady state.

\bigskip
\bigskip

\noindent
{\bf Acknowledgment.} Both authors acknowledge the hospitality of Chris John Budd at the Department of Mathematical Sciences of the Bath Institute for Complex System at the University of Bath, Bath BA2 7AY, United Kingdom, where this research was carried out.
This work was supported by the Italian GNCS of INDAM and by the University of Messina.

\end{document}